\documentclass[runningheads]{llncs}
\usepackage{graphicx}
\usepackage{amsmath}
\usepackage{amssymb}
\usepackage{graphicx}
\usepackage{multirow}
\usepackage[colorinlistoftodos]{todonotes}
\usepackage[colorlinks=true, allcolors=blue]{hyperref}
\usepackage{tikzsymbols}
\usepackage{subfig}
\usepackage{stfloats}
\usepackage{boldline}
\usepackage{comment}
\usepackage{cite}
\usepackage{color}

\begin{document}
\newcommand{\tom}[1]{\textcolor{red}{#1}}
\newcommand{\veronika}[1]{\textcolor{orange}{#1}}
\newsavebox\CBox
\def\textBF#1{\sbox\CBox{#1}\resizebox{\wd\CBox}{\ht\CBox}{\textbf{#1}}}
\titlerunning{Predicting Scores of MI Segmentation Methods with Meta-Learning}
\title{Predicting Scores of Medical Imaging Segmentation Methods with Meta-Learning}


%
%
\author{Tom van Sonsbeek\and Veronika Cheplygina}


%
%
\institute{Eindhoven University of Technology, The Netherlands\\
\email{t.j.v.sonsbeek@gmail.com, v.cheplygina@tue.nl}\\
}
%
\maketitle              
\begin{abstract}

Deep learning has led to state-of-the-art results for many medical imaging tasks, such as segmentation of different anatomical structures. With the increased numbers of deep learning publications and openly available code, the approach to choosing a model for a new task becomes more complicated, while time and (computational) resources are limited. A possible solution to choosing a model efficiently is meta-learning, a learning method in which prior performance of a model is used to predict the performance for new tasks. We investigate meta-learning for segmentation across ten datasets of different organs and modalities. We propose four ways to represent each dataset by meta-features: one based on statistical features of the images and three are based on deep learning features. We use support vector regression and deep neural networks to learn the relationship between the meta-features and prior model performance. On three external test datasets these methods give Dice scores within 0.10 of the true performance. These results demonstrate the potential of meta-learning in medical imaging. 

\keywords{Meta-learning  \and segmentation \and feature extraction.}
\end{abstract}


\section{Introduction}

Deep learning algorithms have become state-of-the-art methods in numerous medical image analysis tasks \cite{LITJENS201760} and have shown to outperform experts on many tasks\cite{liu2019comparison}. Different models have been developed, such as various extensions of convolutional neural networks, recurrent neural networks, and generative adversarial networks, with their respective strengths. Since no model can perform the best on all problems \cite{NFL2002}, for new datasets still new models are being developed. This has led to a dramatic increase of literature: every day around 30 new papers in this field of study are published. There is therefore a need to generalize from all this experience, when selecting a good model for a new medical imaging problem. We propose to do this using meta-learning, a learning method in which prior performance of a model is used to predict the performance for new tasks\cite{lemke_metalearning:_2015, metareview, Vilalta2002}. 

Currently learning from previous experience is largely done through transfer learning. It is possible to outperform training from scratch by transferring model weights or re-using a model for a different task, as shown by Tajbakhsh et al. \cite{tajbakhsh2016convolutional} and Shin et al. \cite{shin2016deep}. However, the quality of that newly created model can only be assessed after training and evaluation, which is costly, both in terms of time and resources. Meta-learning, which has mainly been studied in machine learning field, offers a potential solution. However, it is largely unknown in medical imaging. A Google Scholar search\footnote{Search done in March 2020} on ``medical imaging'' and ``deep learning'' shows that only 270 papers - less than 1\% - are also related to meta-learning. This is possibly due to the complexity of the data or limited differences between datasets. Another reason is data availability - although datasets and models performances are increasingly being shared online, it is only a recent development that challenges focus on multiple applications, for example~\cite{simpson2019large}. 

We propose to use meta-learning to predict segmentation scores across ten datasets of different organs and modalities. We propose four ways to represent each dataset by meta-features: one based on statistical features of the images and three are based on deep learning features. We use support vector regression and a deep neural network to learn the relationship between the meta-features and prior model performance.  

\subsection{Related work}

A common application of meta-learning in computer vision is prediction ranking between methods \cite{finn2017one, ranking2009, rossi2014metastream}. An extension to this is predicting the result of a model. Guerra et al. predicted the outcome of multi-layer perceptron networks using regression models as a meta-learner. This was achieved using compressed representations of datasets, called meta-features. The regression model learns a relationship between the metafeature and prior performance information. Similar approaches were followed by Doan et al. \cite{doan2017predicting} for predicted running time of algorithms and  Soares et al. for predicting the outcome of clustering algorithms \cite{ranking2009}. Gomes et al. \cite{gomes2012combining} and Soares et al. \cite{soares2004meta} used meta-learning to predict parameter settings for support vector machines.

Meta-learning has been applied to a small number of problems in medical imaging. Campos et al. used meta-learning to predict segmentation scores of photos of wounds~\cite{campos2016meta}. While this was not a typical medical dataset, it showed the possibilities of meta-learning in the medical domain. Hu et al. created a meta-learning method which initialised weights for finetuning of classification methods in medical imaging \cite{hu2018meta}, thus reducing the need for data. Cheplygina et al. characterized medical image segmentation problems in meta-feature space defined by performances of classical classifiers~\cite{cheplygina2017exploring}, but only predicted whether datasets originate from the same source. To the best of our knowledge no attempts have yet been made to recommend models by predicting the performance of typical segmentation problems in medical imaging by using meta-learning.

\section{Methods}

We assume we are given a collection of datasets $\{D_i\}^{N}_{1}$, for example segmentation tasks of different organs and modalities. We also assume we have a collection of models $\{U_j\}^{M}_{1}$, for example various U-Net-type architectures, and results $\{y_{ij}\}^{N,M}_{i=1,j=1}$ of these models on the datasets. The challenge is, given a previously unseen dataset $D_{N+1}$, to predict the model scores $\{y_{N+1,j}\}^{M}_{j=1}$.    

\begin{figure}
    \centering
    \includegraphics[width=\textwidth]{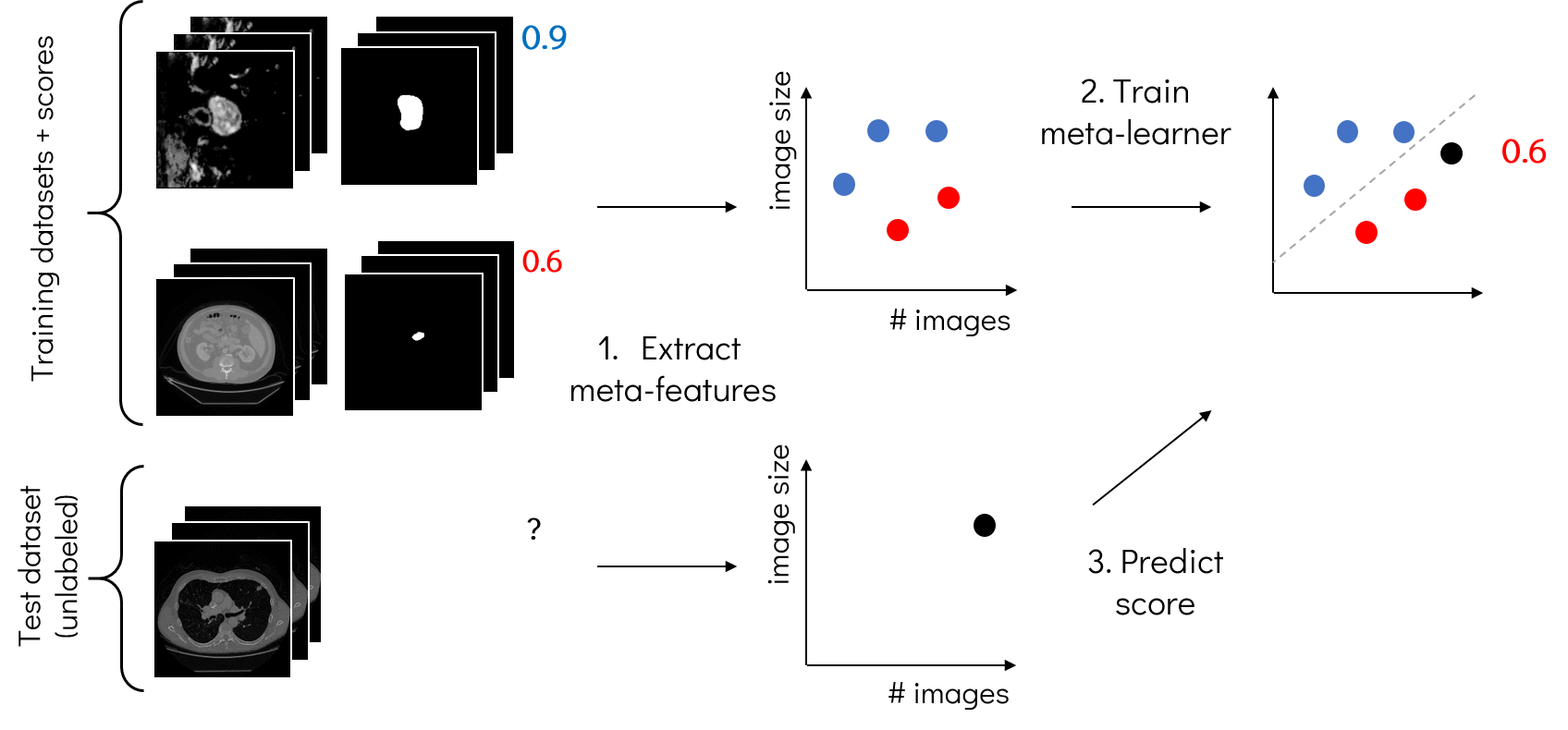}
    \caption{Proposed meta-learning method: training is done on extracted meta-features and meta-labels (scores of segmentation algorithms). At test time, the trained meta-learner can predict scores for a previously unseen dataset. For simplicity here we illustrate a classification problem, but throughout the paper regression is used.}
    \label{fig:metalearner}
\end{figure}

The overall method is illustrated in Fig.~\ref{fig:metalearner}. First a meta-feature extractor $f$ summarizes a subset $s$ of each dataset, such that $\mathbf{x}_i = f(D_i)$ is a $q$-dimensional feature vector. Using a fixed subset size ensures invariance to dataset size. Then for the $j$-th model, a classifier $g_{j}$ is trained on the meta-feature vectors $\mathbf{x}_i$ and the meta-labels $y_{ij}$. Predicting the model's score for an unseen dataset is done via $\hat{y} = g_{j}(f(D_{N+1}))$

\subsection{Meta-feature extraction}

We investigate three broad types of meta-features, described in more detail below:
\begin{itemize}
    \item Classical meta-features, similar to meta-features used in prior work.
    \item Deep learning based meta-features with three different architectures: VGG16, ResNet50, MobileNetV1.
    \item Task-specific meta-features, which provide context about a given segmentation problem; added to both the classical and deep learning meta-features. 
\end{itemize}

\subsubsection{Classical meta-features}
We used meta-features from classical (non-imaging) applications of meta-learning. \cite{10.1007/3-540-36182-0_14, metareview, campos2016meta}. Typical examples are mean pixel value, dataset correlation and entropy. These meta-features will be referred to as classical meta-features. A selection requirement for each classical meta-feature is that it should be visually different between datasets. To check whether this criteria has been met a visual inspection of each meta feature will be done. This selection led to meta-features $\mathbf{x}_{i} = f_{CLAS}(D_i)$ with $\mathbf{x}_{i} \in \mathbb{R}^{33} $ is available in the Supplementary Material. 

\subsubsection{Deep learning meta-features}
\begin{figure}[!]
    \centering
    \includegraphics[width =\textwidth]{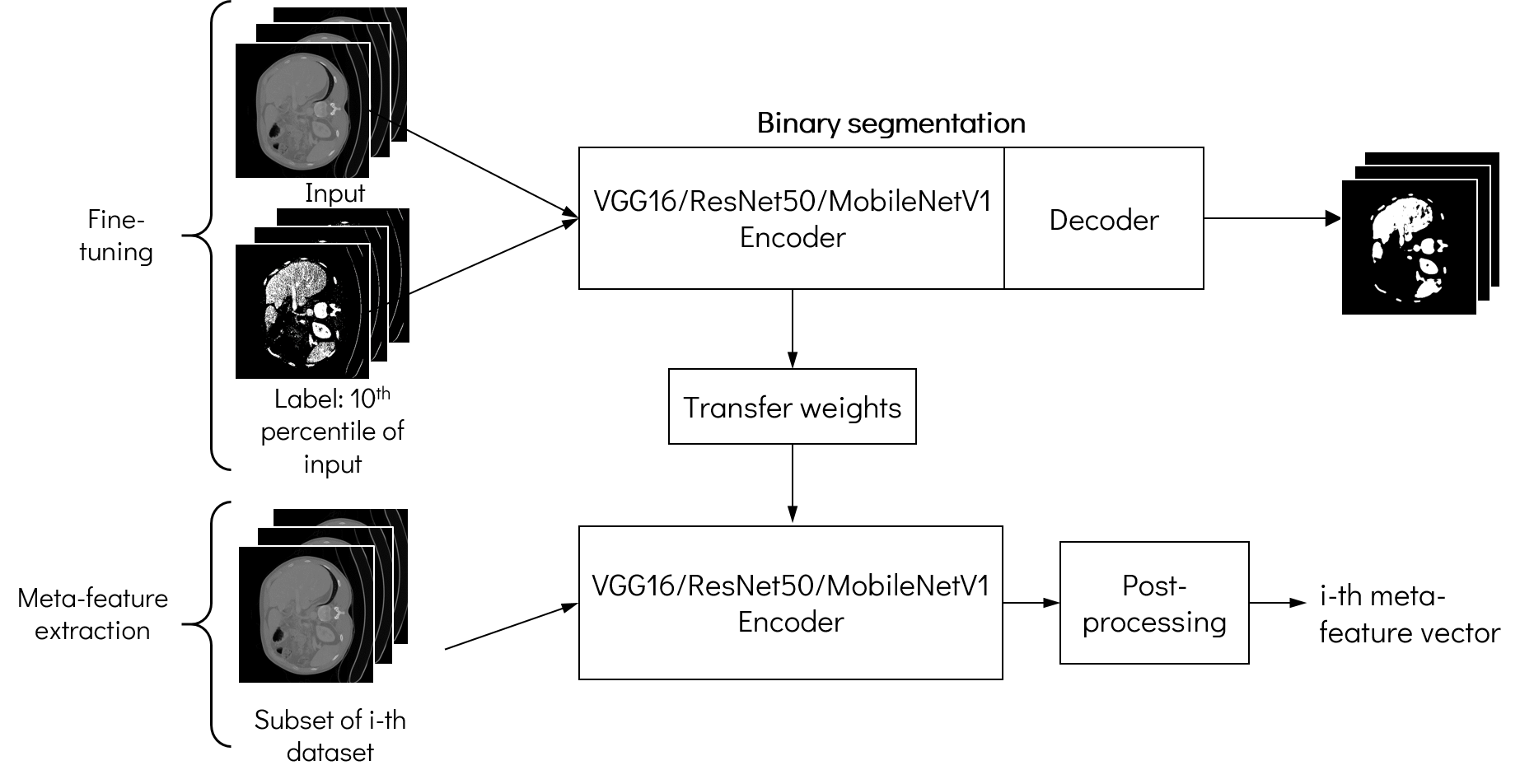}
    \caption{Deep learning based meta-feature extraction.}
    \label{fig:femet}
\end{figure}

Deep learning is successful for feature extraction at an image level, we therefore also investigated how it can be applied to extract features at a dataset level. Initial attempts using networks pretrained on ImageNet without fine-tuning did not lead to distinctive features for datasets of the same modality. Therefore we added a fine-tuning step with a U-Net-like encoder-decoder network, where the encoder is the feature extractor and the decoder is the original U-Net \cite{DBLP:journals/corr/RonnebergerFB15}. This encoder-decoder network works as a binary segmentation network that fine-tunes the weights of the feature extractor (\autoref{fig:femet}). 

An important property of the fine-tuning step is that labels from the test data cannot be used, since the meta-learning method should be able to generalize to datasets without ground-truth segmentations. Instead, we introduce an auxiliary task by thresholding the voxels intensities at the 10-th percentile to create rough segmentation masks. These masks are not accurate in terms of segmentation, but provide a good enough estimation of the image structure.      

Using this strategy, we fine-tune three different models, pretrained on ImageNet: VGG16, ResNet50, and MobileNetV1. An intermediate step of these meta-features $x_{i} = f_{DL}(D_i)$) with $\mathbf{x}_{i} \in \mathbb{R}^{(z, 7, 7)} $, $z \in \{512, 2048, 1024\}$ consist of the output of the last layers of these models, averaged over the number of images on which the meta-feature is computed. The $7\times7$ feature maps are binarized yielding final meta-features $\mathbf{x}_{i} \in \mathbb{R}^{z}$. This binarization is done by thresholding using computing feature map correlation across datasets during training time, with an empirically determined threshold: $\alpha = 0.80$. To improve the meta-feature quality univariate feature selection is applied, where the optimal selection threshold is based on SVM classifier weights during training time. 

\subsubsection{Task-specific meta-features}

Additionally, we supplement both classical and deep learning meta-features above with task-specific meta-features which capture basic properties of the datasets, such as the modality. These meta-features could be queried from the user, or detected with simple classifiers. For simplicity, here we have set the following features between 0 and 1 based on exploratory analysis of each dataset: imaging modality, whether the segmentation is location-dependent, how sphere-shaped is the segmentation, relative size of the segmentation, and presence of multiple segmentation objects. 

\subsection{Meta-learner}

The meta-learner is a model which relates the meta-features to segmentation scores. For this meta-learner two methods are used. The first method uses support vector regression (SVR)\cite{scikit-learn}, a common regression method in machine learning. Default parameter settings are used. The second method uses a three layer deep fully connected multi-layer perceptron network (DNN), with ReLu activated hidden layers of sizes 50 and 30.  A dropout rate of 50\% is used. The last layer is sigmoid activated to result in the final prediction.

\subsection{Evaluation}
The mean absolute error (MAE) is used as the scoring function. This is a common metric in similar meta-learning methods which use regression methods \cite{gomes2012combining, prudencio2004meta}, see \autoref{eq:mae}: 
\begin{equation}
    MAE = {\frac{1}{n}\sum_{i=1}^{n}|y_{i}-\hat{y}_{i}|}.
    \label{eq:mae}
\end{equation}
Furthermore to assess whether the meta-learner is not simply predicting the same score for every dataset (prediction towards the mean), we use the normalized mean absolute error (NMAE) \cite{soares2004meta}. A NMAE (\autoref{eq:nmae}) score of 1 means performance is equal to always predicting the mean performance of the training datasets. NMAE values higher than 1 mean that the meta-learners performs worse than the mean performance prediction. Values lower than 1 are desired. Using this metric meta-learners on different problems can be compared.
\begin{equation}
    NMAE = \frac{\sum_{i=1}^{n}|y_{i}-\hat{y}_{i}|}{\sum_{i=1}^{n}|y_{i}-\bar{y}_{i}|}.
    \label{eq:nmae}
\end{equation}

\section{Experiments}

For the first part of the experiments, we use data from the Medical Segmentation Decathlon (MSD) challenge~\cite{simpson2019large}. The goal of this challenge was to develop a model which could, after a fine-tuning step, segment several distinct segmentation problems. Ten datasets with varying anatomical regions and imaging modalities (CT and MR) were included. We used these datasets, and performances of challenge participants, for our meta-learning method. Additionally, we used performances of the winning participant \cite{nnUnet} on three public datasets, as held-out test datasets: LiTS (liver CT) \cite{Lits}, ACDC (heart MR) \cite{acdc}, CHAOS (liver CT) \cite{kavur2020chaos}. 

\subsection{Meta-feature generation}

\begin{figure}[]
\centering
\includegraphics[width=10cm]{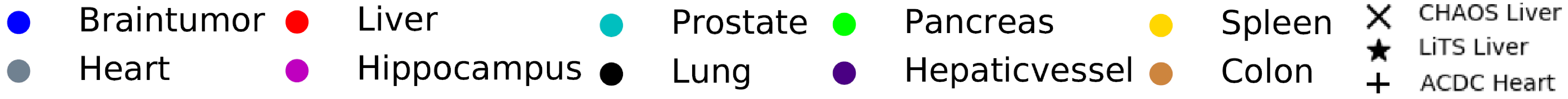}\\

\subfloat[][Classical]{\includegraphics[width=3cm]{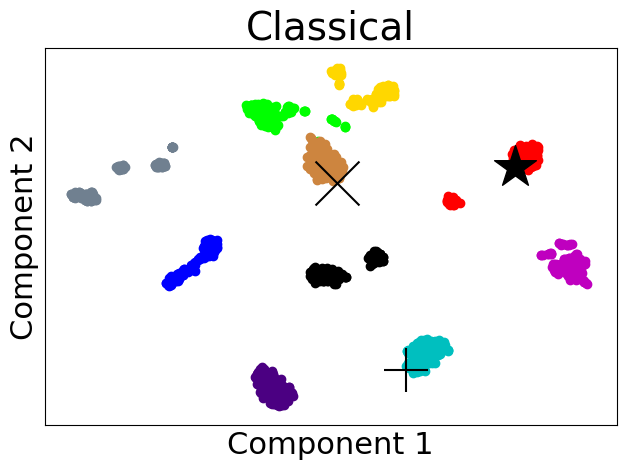} }
\subfloat[][VGG16]{\includegraphics[width=3cm]{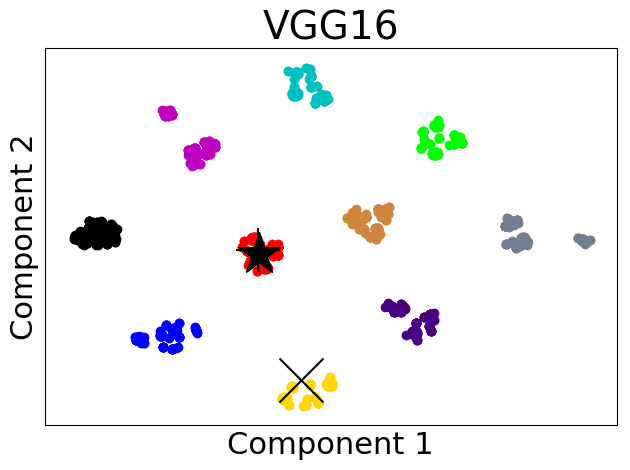} }
\subfloat[][ResNet50]{\includegraphics[width=3cm]{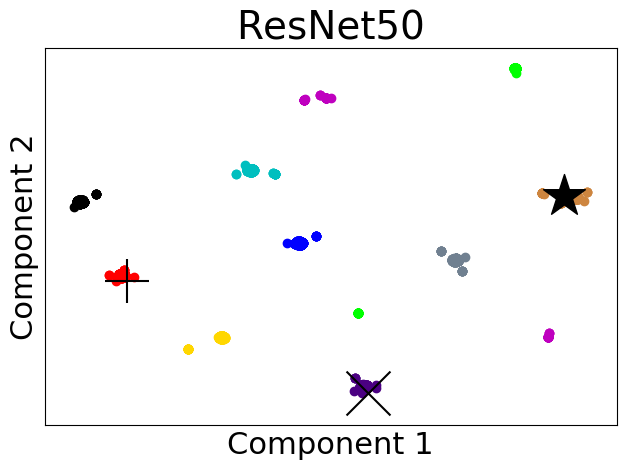} }
\subfloat[][MobileNetV1]{\includegraphics[width=3cm]{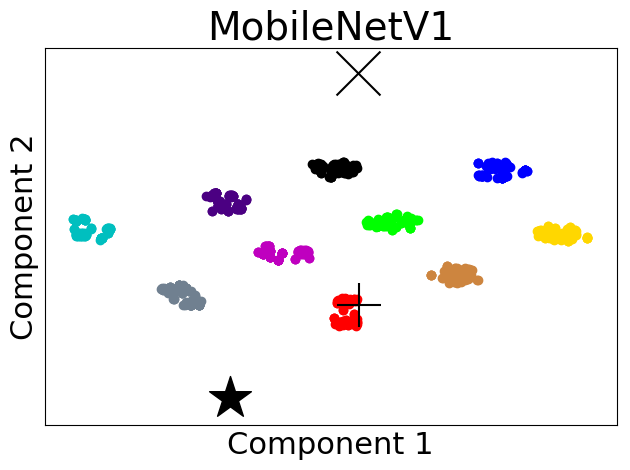} }\\

\caption{t-SNE embeddings of meta-features from MSD datasets and test datasets.}%
\label{fig:tsne}%

\end{figure}

A meta-feature vector is based on a subset of $s = 20$ images, sampled from the dataset. A total of 100 subsets, and thus meta-feature vectors are sampled from each dataset. We first examined the quality of the different meta-feature types using the t-stochastic nearest neighbor (t-SNE) embeddings for the MSD datasets. The results are shown in Fig.~\ref{fig:tsne}. The embeddings show that all meta-features are able to separate the datasets well, but the deep learning meta-features provide more well-defined separation. 

\subsection{Cross-validation MSD}

We then performed experiments with the MSD datasets to determine the performance of different meta-features and meta-learners. We used cross-validation with 7 datasets for training and 3 datasets for testing. 

We show the performances of the different combinations in Table~\ref{tab:totalcross}. Consistent with the t-SNE embeddings, we see that the deep learning meta-features lead to lower errors than the classical meta-features. Out of the deep learning features, ResNet50 leads to the best results. Furthermore, we see that the SVR and DNN meta-learners perform on par with each other. In general, the lower the intra-variability of segmentation scores within a dataset, the higher the predictive accuracy. Results for individual datasets and challenge participants  can be found in the Supplementary Material.  

\captionsetup{width=10cm}
\begin{table}[!t]
\centering
\resizebox{8cm}{!}{%
\begin{tabular}{l|ll|ll}
                                                  \multirow{2}{*}{$\downarrow$ Feature extractor } & \multicolumn{2}{c|}{MAE} & \multicolumn{2}{c}{NMAE}\\ \cline{2-5}
 & \multicolumn{1}{c}{SVR}                    & \multicolumn{1}{c|}{DNN}                    & \multicolumn{1}{c}{SVR}                    & \multicolumn{1}{c}{DNN} \\ \hline
\multicolumn{1}{l|} {Task-specific only} & 0.22 $\pm$ 0.13          & 0.26 $\pm$ 0.12          & 1.13 $\pm$ 0.66           & 1.33 $\pm$ 0.61\\ \hline
\multicolumn{1}{l|} {Statistical} & 0.21 $\pm$ 0.08          & 0.24 $\pm$ 0.10          & 1.07 $\pm$ 0.46           & 1.19 $\pm$ 0.51\\ \hline
\multicolumn{1}{l|} {VGG16} & 0.13 $\pm$ 0.09          & 0.14 $\pm$  0.04         & 0.65 $\pm$ 0.46          & 0.70 $\pm$ 0.21\\
\hline
\multicolumn{1}{l|}{ResNet50} & \textBF{0.12 $\pm$ 0.07} & \textBF{0.12 $\pm$ 0.03} & \textBF{0.62 $\pm$ 0.36} & \textBF{0.62 $\pm$ 0.15}\\
\hline
\multicolumn{1}{l|}{MobileNetV1} & 0.15 $\pm$ 0.09          & 0.14 $\pm$ \textBF{0.03}          & 0.76 $\pm$ 0.46          & 0.70 $\pm$ \textBF{0.15}\\
\end{tabular}%
}
\caption{Cross-validation result of SVR and DNN meta-learners on MSD data for four different meta-features. Bold = best result per column.}
\label{tab:totalcross}
\end{table}
\captionsetup{width=\textwidth}

\begin{figure}[!b]
\centering

\includegraphics[width=9cm]{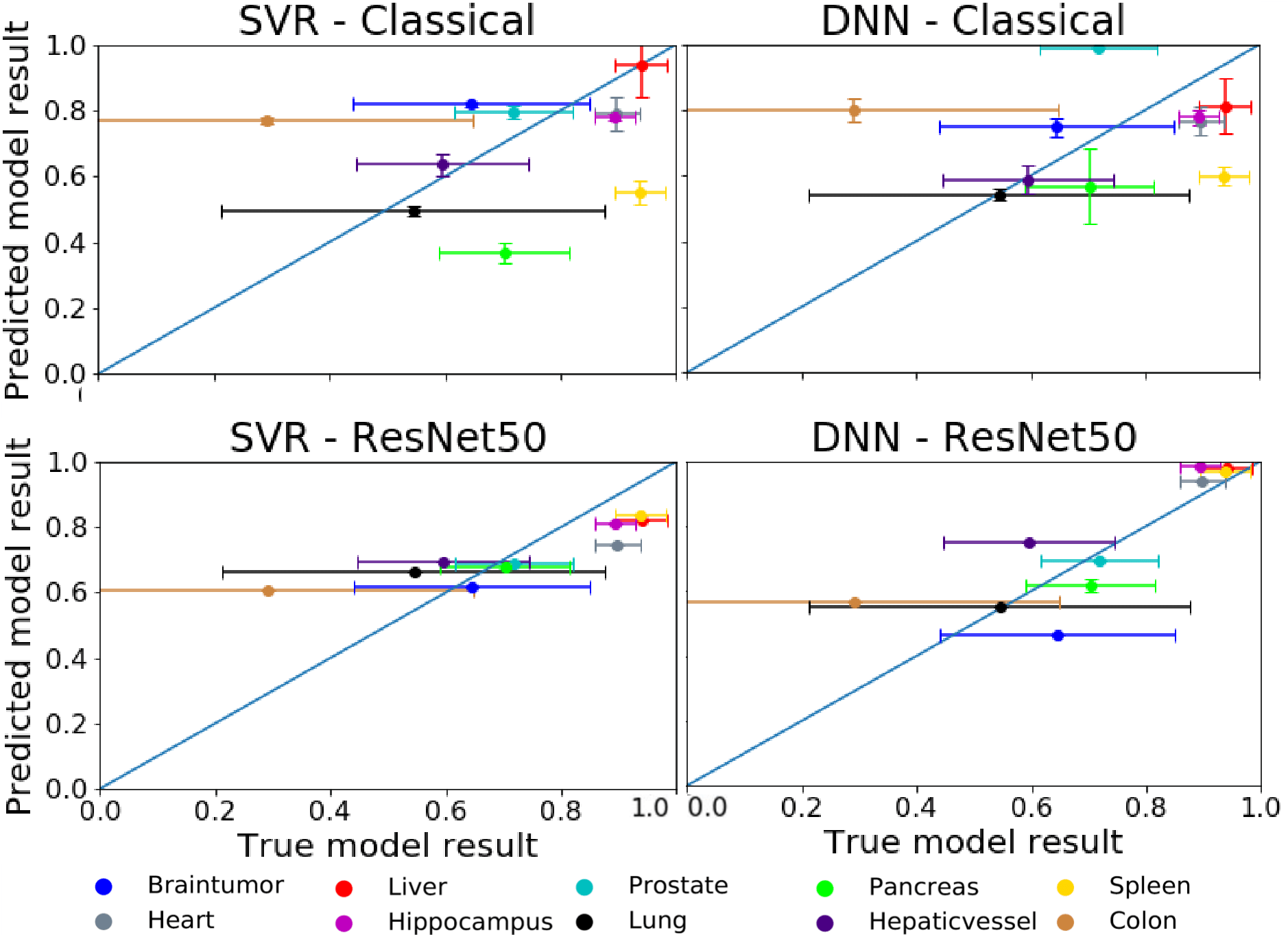}
\caption{Examples of results of Cross-validation on MSD datasets for Classifical and ResNet50 meta-features, and SVR and DNN meta-learners.}%
\label{fig:cor}
\end{figure}

To further examine the behavior of different methods, we plot the predicted Dice scores against the true Dice scores in Fig.~\ref{fig:cor}. Here we can see that the overall correlation is positive, but for some datasets the predictions are better than for others. Datasets with ``average'' scores consistently yield low prediction errors. 


\subsection{Held-out test data}
We then do a similar experiment as before, but instead of cross-validation on 10 datasets, we train the meta-learners on the MSD data, and test them on three held-out datasets. 

The MAE results are shown in Table~\ref{tab:testtotalmae}. Prediction results can be found in the Supplementary Material. Comparing the meta-features, we see that the classical meta-features are best for two out of three datasets when SVR is used, and the three deep learning features are best once when DNN is used. Averaging the results across the datasets, the DNN meta-learner has the lowest error.

\begin{table*}[!t]
\centering
\resizebox{\textwidth/10*8}{!}{%
\begin{tabular}{l|c c|c c|c c|c c|c c}
\multirow{2}{*}{$\downarrow$ Feature extractor} & \multicolumn{2}{c|}{Liver (LiTS)} & \multicolumn{2}{c|}{Heart (ACDC)} & \multicolumn{2}{c|}{Liver (CHAOS)} & \multicolumn{2}{c|}{\textBF{Mean MAE}} & \multicolumn{2}{c}{\textBF{Mean NMAE}} \\ \cline{2-11} 
 & SVR & DNN & SVR & DNN & SVR & DNN & \textBF{SVR} & \textBF{DNN} & \textBF{SVR} & \textBF{DNN} \\ \hline
{Task-specific only}
 & $0.10$ & $0.46$ & $0.17$ & $0.42$ & $0.02$ & $0.44$ & $0.10$ & $0.44$ & $0.61$ & $2.72$ \\ \hline
{Statistical}
 & \textBF{0.09} & $0.14$ & $0.16$ & $0.16$ & \textBF{0.01} & $0.04$ & \textBF{0.09} & 0.11 & \textBF{0.54} & $0.71$ \\ \hline
{VGG16}  & $0.20$ & $0.13$ & $0.22$ & $0.05$ & $0.15$ & \textBF{0.01} & $0.19$ & \textBF{0.06} & $1.19$ & \textBF{0.40} \\ \hline
{ResNet50} & $0.23$ & $0.50$ & $0.23$ & $$\textBF{0.02}$$ & $0.15$ & $0.06$ & $0.20$ & $0.19$ & $1.27$ & $1.21$ \\ \hline
{MobileNetV1} & $0.14$ & \textBF{0.01} & \textBF{0.14} & $0.13$ & $0.07$ & $0.07$ & $0.14$ & $0.07$ & $0.88$ & $0.44$ \\
\end{tabular}%
}
\caption{MAE scores of SVR and DNN meta-learners on test datasets for different types of metafeatures.}
\label{tab:testtotalmae}
\end{table*}


\section{Discussion}

We investigated whether meta-learners can predict the performance of segmentation algorithms, based on various meta-feature representations of datasets. We found that the predicted Dice scores are within a 0.10 of the true results, which is a promising result. While such a method would not help between distinguishing among the top few methods for a particular segmentation problem, it could eliminate some alternatives that are not suitable. 

The proposed study still has some limitations. One issue is that the datasets are quite sparse, with a low number and large differences between datasets. We would recommend including more datasets which share either task and/or modality with the existing datasets. 

Furthermore, our method assumes all segmentation methods under consideration have been tested on all the available datasets. This scenario is still limited to challenges, although a platform where different datasets and models are shared, such as OpenML\cite{OpenML2013}, could be a possibility in the future. Furthermore, meta-learners which can be trained with missing data, could also be investigated.

\section{Conclusion}
Prediction of performance using these meta-features yields promising results. The error margins of the methods are still too large for decision-making based on the outcome of this meta-learning method, but is is clearly shown that prior performance of methods in combination with dataset characteristics is a predictor of performance and can lead to a more efficient way of development.

\bibliography{refs_veronika.bib, references.bib}
\bibliographystyle{splncs04}
\newpage
\appendix

{
\centering
\Large\bfseries
Appendices
\centering
}

\section{Classical metafeatures}
List of 33 classical features used to compose the classical metafeatures.
\setcounter{table}{0}
\renewcommand{\thetable}{A.\arabic{table}}
\label{appendix:stat}
\begin{table}[!h]
\centering
\resizebox{\textwidth/3+6mm}{!}{
\begin{tabular}{l}
\hlineB{4}
\textbf{Classical metafeatures}\\\hlineB{4}
Number of instances\\\hline
Voxel value M\\\hline
Voxel value STD\\\hline
Voxel value CVAR\\\hline
Skew M\\\hline
STD\\\hline
Skew CVAR\\\hline
Kurtosis M\\\hline
Kurtosis STD\\\hline
Kurtosis CVAR\\\hline
Entropy M\\\hline
Entropy STD\\\hline
Entropy CVAR\\\hline
Median M\\\hline
Median STD\\\hline
Mutual information M\\\hline
Mutual information STD\\\hline
Mutual information CVAR\\\hline
Mutual information maximum value\\\hline
Correlation M\\\hline
Correlation STD\\\hline
Correlation CVAR\\\hline
Sparsity M\\\hline
Sparsity STD\\\hline
Sparsity CVAR\\\hline
Slice size M\\\hline
Slice size STD\\\hline
Slice size CVAR\\\hline
Number of slices M\\\hline
Number of slices STD \\\hline
Number of slices CVAR\\\hline
Equivalent number of features\\\hline
Noise signal ratio\\\hline
\end{tabular}}

\captionsetup{width=.8\textwidth}

\caption{Classical metafeatures used in the Support Vector Regression method. M = mean, STD = standard deviation, CVAR = coefficient of variation. }

\end{table}

\clearpage

\section{Full results MSD cross-validation with SVR and DNN meta-learner}
Full results of SVR and DNN meta-learners. Consists of: MAE scores per MSD dataset and MAE scores per MSD challenge participant.
\label{appendix:svrtot}

\setcounter{table}{0}
\renewcommand{\thetable}{B.\arabic{table}}
\begin{table}[htp]
\centering
\resizebox{10.5cm}{!}{%
\begin{tabular}{|ll|l|l|l|l|l|l|l|l|}
\hline
\multicolumn{2}{|c|}{\multirow{2}{*}{MSD datasets $\downarrow$}} & \multicolumn{8}{c|}{Mean absolute error $\downarrow$} \\ \cline{3-10} 
\multicolumn{2}{|c|}{} & \multicolumn{2}{c|}{Classical} & \multicolumn{2}{c|}{VGG16} & \multicolumn{2}{c|}{ResNet50} & \multicolumn{2}{c|}{MobileNetV1} \\ \cline{3-10} 
\multicolumn{2}{|l|}{Meta-learner $\rightarrow$} & \multicolumn{1}{c|}{SVR} & DNN & \multicolumn{1}{c|}{SVR} & DNN & \multicolumn{1}{c|}{SVR} & DNN & \multicolumn{1}{c|}{SVR} & DNN \\ \hline
1  & Braintumor     & $0.17\pm0.05$ & $0.06\pm0.06$ & $0.07\pm0.01$ & $0.09\pm0.01$ & $0.08\pm0.01$ & $0.22\pm0.06$ & $0.14\pm0.01$ & $0.22\pm0.06$ \\ \hline
2  & Heart          & $0.07\pm0.01$ & $0.18\pm0.05$ & $0.06\pm0.05$ & $0.09\pm0.01$ & $0.14\pm0.06$ & $0.05\pm0.01$ & $0.15\pm0.07$ & $0.11\pm0.02$ \\ \hline
3  & Liver          & $0.29\pm0.15$ & $0.55\pm0.25$ & $0.08\pm0.17$ & $0.03\pm0.00$ & $0.11\pm0.10$ & $0.04\pm0.00$ & $0.28\pm0.15$ & $0.24\pm0.06$ \\ \hline
4  & Hippocampus    & $0.07\pm0.01$ & $0.09\pm0.01$ & $0.04\pm0.07$ & $0.07\pm0.01$ & $0.08\pm0.07$ & $0.10\pm0.01$ & $0.06\pm0.07$ & $0.07\pm0.01$ \\ \hline
5  & Prostate       & $0.14\pm0.03$ & $0.27\pm0.09$ & $0.09\pm0.00$ & $0.21\pm0.05$ & $0.05\pm0.00$ & $0.07\pm0.01$ & $0.04\pm0.00$ & $0.05\pm0.01$ \\ \hline
6  & Lung           & $0.09\pm0.01$ & $0.09\pm0.01$ & $0.21\pm0.11$ & $0.23\pm0.06$ & $0.17\pm0.07$ & $0.10\pm0.02$ & $0.09\pm0.09$ & $0.09\pm0.02$ \\ \hline
7  & Pancreas       & $0.19\pm0.03$ & $0.07\pm0.01$ & $0.10\pm0.01$ & $0.09\pm0.01$ & $0.06\pm0.01$ & $0.09\pm0.01$ & $0.05\pm0.01$ & $0.16\pm0.04$ \\ \hline
8  & Hepatic vessel & $0.12\pm0.01$ & $0.10\pm0.05$ & $0.26\pm0.03$ & $0.21\pm0.06$ & $0.13\pm0.04$ & $0.21\pm0.05$ & $0.13\pm0.06$ & $0.09\pm0.02$ \\ \hline
9  & Spleen         & $0.46\pm0.18$ & $0.40\pm0.18$ & $0.08\pm0.14$ & $0.03\pm0.00$ & $0.10\pm0.11$ & $0.03\pm0.00$ & $0.21\pm0.10$ & $0.12\pm0.02$ \\ \hline
10 & Colon          & $0.54\pm0.23$ & $0.56\pm0.32$ & $0.31\pm0.28$ & $0.34\pm0.13$ & $0.31\pm0.25$ & $0.33\pm0.15$ & $0.36\pm0.30$ & $0.25\pm0.08$ \\ \hline
\textbf{} & \textbf{Total} & \multicolumn{1}{l|}{\textbf{$0.21\pm0.08$}} & \textbf{$0.24\pm0.09$} & \multicolumn{1}{l|}{\textbf{$0.13\pm0.09$}} & \textbf{$0.14\pm0.04$} & \multicolumn{1}{l|}{\textbf{$0.12\pm0.07$}} & \textbf{$0.12\pm0.03$} & \multicolumn{1}{l|}{\textbf{$0.15\pm0.09$}} & \textbf{$0.14\pm0.03$} \\ \hline
\end{tabular}%
}
\caption{Total MAE results of cross-validation on MSD datasets per MSD dataset using different types of meta-features and SVR and DNN meta-learner}
\label{tab:maetotmsvr}
\end{table}
\begin{table}[htp]
\centering
\resizebox{10.5cm}{!}{%
\begin{tabular}{|l|l|l|l|l|l|l|l|l|}
\hline
\multirow{2}{*}{Participants $\downarrow$} & \multicolumn{8}{c|}{Mean absolute error $\downarrow$} \\ \cline{2-9} 
 & \multicolumn{2}{c|}{Clasical} & \multicolumn{2}{c|}{VGG16} & \multicolumn{2}{c|}{ResNet50} & \multicolumn{2}{c|}{MobileNetV1} \\ \cline{2-9} 
Meta-learner $\rightarrow$ & SVR & DNN & SVR & DNN & SVR & DNN & SVR & DNN \\ \hline
Participant 1  & $0.25\pm0.08$ & $0.29\pm0.29$ & $0.08\pm0.11$ & $0.11\pm0.02$ & $0.18\pm0.10$ & $0.20\pm0.08$ & $0.19\pm0.12$ & $0.21\pm0.06$ \\ \hline
Participant 2  & $0.18\pm0.06$ & $0.19\pm0.19$ & $0.11\pm0.05$ & $0.11\pm0.02$ & $0.10\pm0.05$ & $0.12\pm0.03$ & $0.13\pm0.05$ & $0.13\pm0.02$ \\ \hline
Participant 3  & $0.20\pm0.05$ & $0.21\pm0.07$ & $0.12\pm0.06$ & $0.10\pm0.02$ & $0.11\pm0.05$ & $0.11\pm0.02$ & $0.13\pm0.06$ & $0.13\pm0.03$ \\ \hline
Participant 4  & $0.27\pm0.09$ & $0.24\pm0.10$ & $0.11\pm0.14$ & $0.14\pm0.03$ & $0.14\pm0.12$ & $0.17\pm0.04$ & $0.18\pm0.14$ & $0.13\pm0.03$ \\ \hline
Participant 5  & $0.19\pm0.70$ & $0.21\pm0.05$ & $0.12\pm0.06$ & $0.12\pm0.03$ & $0.11\pm0.05$ & $0.09\pm0.02$ & $0.12\pm0.06$ & $0.11\pm0.02$ \\ \hline
Participant 6  & $0.24\pm0.10$ & $0.25\pm0.11$ & $0.13\pm0.09$ & $0.15\pm0.03$ & $0.14\pm0.08$ & $0.11\pm0.02$ & $0.19\pm0.10$ & $0.18\pm0.04$ \\ \hline
Participant 7  & $0.21\pm0.07$ & $0.29\pm0.09$ & $0.16\pm0.12$ & $0.20\pm0.05$ & $0.15\pm0.09$ & $0.12\pm0.03$ & $0.15\pm0.11$ & $0.13\pm0.02$ \\ \hline
Participant 8  & $0.15\pm0.03$ & $0.16\pm0.02$ & $0.07\pm0.03$ & $0.07\pm0.01$ & $0.09\pm0.03$ & $0.07\pm0.01$ & $0.12\pm0.03$ & $0.11\pm0.02$ \\ \hline
Participant 9  & $0.24\pm0.10$ & $0.23\pm0.08$ & $0.14\pm0.08$ & $0.15\pm0.03$ & $0.12\pm0.07$ & $0.08\pm0.01$ & $0.22\pm0.08$ & $0.15\pm0.04$ \\ \hline
Participant 10 & $0.21\pm0.08$ & $0.27\pm0.10$ & $0.17\pm0.10$ & $0.19\pm0.06$ & $0.14\pm0.08$ & $0.16\pm0.06$ & $0.13\pm0.10$ & $0.12\pm0.02$ \\ \hline
Participant 11 & $0.24\pm0.09$ & $0.26\pm0.14$ & $0.17\pm0.09$ & $0.18\pm0.07$ & $0.14\pm0.08$ & $0.16\pm0.06$ & $0.15\pm0.1$  & $0.18\pm0.05$ \\ \hline
Participant 12 & $0.26\pm0.10$ & $0.26\pm0.15$ & $0.17\pm0.11$ & $0.20\pm0.06$ & $0.16\pm0.09$ & $0.16\pm0.04$ & $0.17\pm0.11$ & $0.15\pm0.04$ \\ \hline
Participant 13 & $0.18\pm0.06$ & $0.23\pm0.60$ & $0.11\pm0.06$ & $0.11\pm0.02$ & $0.01\pm0.05$ & $0.12\pm0.03$ & $0.14\pm0.06$ & $0.15\pm0.03$ \\ \hline
Participant 14 & $0.15\pm0.04$ & $0.18\pm0.03$ & $0.08\pm0.04$ & $0.09\pm0.01$ & $0.09\pm0.03$ & $0.08\pm0.01$ & $0.10\pm0.04$ & $0.10\pm0.02$ \\ \hline
Participant 15 & $0.17\pm0.04$ & $0.19\pm0.04$ & $0.09\pm0.05$ & $0.08\pm0.01$ & $0.10\pm0.04$ & $0.10\pm0.01$ & $0.12\pm0.05$ & $0.10\pm0.02$ \\ \hline
Participant 16 & $0.22\pm0.09$ & $0.26\pm0.12$ & $0.15\pm0.08$ & $0.14\pm0.04$ & $0.12\pm0.07$ & $0.12\pm0.03$ & $0.14\pm0.09$ & $0.13\pm0.03$ \\ \hline
Participant 17 & $0.20\pm0.08$ & $0.24\pm0.09$ & $0.13\pm0.08$ & $0.13\pm0.04$ & $0.12\pm0.07$ & $0.11\pm0.03$ & $0.14\pm0.08$ & $0.16\pm0.04$ \\ \hline
Participant 18 & $0.28\pm0.12$ & $0.31\pm0.15$ & $0.21\pm0.15$ & $0.25\pm0.07$ & $0.15\pm0.11$ & $0.22\pm0.07$ & $0.19\pm0.13$ & $0.22\pm0.06$ \\ \hline
Participant 19 & $0.22\pm0.08$ & $0.25\pm0.09$ & $0.16\pm0.09$ & $0.15\pm0.05$ & $0.12\pm0.08$ & $0.11\pm0.02$ & $0.14\pm0.09$ & $0.13\pm0.03$ \\ \hline
\textbf{Total} & \textbf{$0.21\pm0.08$} & \textbf{$0.24\pm0.10$} & \textbf{$0.13\pm0.09$} & \textbf{$0.14\pm0.04$} & \textbf{$0.13\pm0.07$} & \textbf{$0.13\pm0.03$} & \textbf{$0.15\pm0.09$} & \textbf{$0.14\pm0.03$} \\ \hline
\end{tabular}%
}
\caption{Total MAE results of cross-validation on MSD datasets per MSD participant using different types of meta-features and SVR and DNN meta-learners}
\label{tab:maetotsvr}
\end{table}
\clearpage

\section{Full prediction results independent test datasets}
Prediction results of SVR and DNN meta-learners on external test datasets (LiTS, ACDC and CHAOS) using different types of meta-features.
\setcounter{table}{0}
\renewcommand{\thetable}{C.\arabic{table}}
\begin{table}[!h]
\centering
\resizebox{8cm}{!}{%
\begin{tabular}{|l|c|c|c|c|c|c|}
\hline
 & \multicolumn{2}{l|}{Liver (LiTS)} & \multicolumn{2}{l|}{Heart (ACDC)} & \multicolumn{2}{l|}{Liver (CHAOS)} \\ \cline{2-7}
True model result $\xrightarrow{}$ & \multicolumn{2}{c|}{0.96} & \multicolumn{2}{c|}{0.96} & \multicolumn{2}{c|}{0.89} \\ \cline{2-7}
$\downarrow$Meta-feature extractor& SVR & DNN & SVR & DNN & SVR & DNN \\ \hline
Classical & 0.87 & 0.82 & 0.80 & 0.8 & 0.90 & 0.85 \\ \cline{1-7}
VGG16 & 0.76 & 0.83 & 0.74 & 0.91 & 0.74 & 0.9 \\ \cline{1-7} 

ResNet50 & 0.73 & 0.46 & 0.73 & 0.94 & 0.74 & 0.95 \\ \cline{1-7} 

MobileNetV1& 0.82 & 0.97 & 0.75 & 0.83 & 0.82 & 0.96 \\ \hline 

\end{tabular}%
}
\caption{Prediction result of SVR and DNN meta-learners on test datasets for different types of meta-features.}
\label{tab:testtrueresult}
\end{table}

\end{document}